\documentclass[aps,prd,twocolumn,showkeys,floats]{revtex4}
\usepackage{graphics,epsfig,color} % figures en eps
\usepackage{amsmath,amssymb}
\usepackage[ansinew]{inputenc}

\begin{document}
%%-----------------------------
%%      the top matter
%%-----------------------------
\title{Explicit form of the Scalar-Tensor metric to be used for propagation of light in the Solar system in continuity of the GR IAU2000 metric}
\author{Olivier Minazzoli\footnote{also at ICRAnet, University of Nice Sophia Antipolis, 28 Avenue Valrose, 06103 Nice, France} and Bertrand Chauvineau}
\affiliation{UNS, OCA-ARTEMIS UMR 6162, Observatoire de la C\^ote d'Azur, Avenue Copernic, 06130 Grasse, France}
\begin{abstract}
The metric recommanded by the IAU2000 resolutions allows propagation of light
calculations at the $c^{-3}$\ level in the general relativity framework.\ In
a recent paper \cite{MCPRD09}, motivated by forthcoming space experiments
involving propagation of light in the Solar System (ASTROD, GAIA, LATOR, ODYSSEY, SAGAS, SIM, TIPO, ...), we have proposed an
extention of the IAU metric equations at the $c^{-4}$\ level. This has been made in the
general relativity framework. However, scalar-tensor theories may induce
corrections numerically comparable to the $c^{-4}$ general relativistic
terms.\ Accordingly, one proposes in this paper an extension of \cite{MCPRD09}
to the scalar-tensor case. The case of a strongly hierarchized system (such as the Solar system) is emphasized. In this case, an explicit metric solution is proposed.

\end{abstract}
\maketitle

%______________________________________________________________________
\section{Introduction}

Forthcoming space missions and missions in project (like ASTROD, TIPO or LATOR, see \cite{MCPRD09}) will require distance measurements at millimetric level in the Solar System.\ This corresponds to time transfer at the precision $%
10^{-11}\;s$. As argued in \cite{MCPRD09}, this requires a complete Solar
System metric at the $c^{-4}$\ level, in order to describe the laser links
involved in such experiments. This has been proposed in the framework of the
General Relativity (GR) theory in \cite{MCPRD09}, leading to an appropriate
extention of the metric recommanded by the IAU2000 resolution \cite{SKPetAJ03}.

The relative amplitude of the relativistic effects is of the order of $\epsilon
=GM/rc^{2}$, where $M$ and $r$ are some characteristic mass and distance.
The so-called first order terms are of order $\epsilon $, the second order
terms of order $\epsilon ^{2}$. In the inner Solar System, $\epsilon $ is
typically of the order $10^{-8}$, and can be sensitively greater ($10^{-7}$)
for photons entering well inside Mercury's orbit, and can even be as large
as $10^{-6}$ for photons grazzing the Sun. On the other hand, GR faces a lot
of difficulties both at the solar system level (Pioneer anomaly, fly-by
anomaly) and at the cosmological level (universe expansion accelarate).\
Hence a new surge of interest in alternative gravity theories.\ Among these, the scalar-tensor (ST) theories deserve a particular interest,
since the gravitational sector of fundamental theories, like string or Kaluza-Klein theories, turns out to be described by a metric tensor plus
a scalar field \cite{FMbook03,Fbook04}  (Brans-Dicke or not \cite{CBPrD07}).

Alternative gravity metrics diverge from the GR one at the $c^{-2}$ level.\
This divergence is quantified by the Post-Newtonian (PN) $\gamma $
factor ($=1$ in GR) entering the $c^{-2}$ term in the space-space components
of the metric tensor. From the present observations, $\left|
\gamma -1\right| $ can at best reach values of the order of $10^{-5}$
\cite{WLivRev06}.\ Besides, some theoretical considerations strongly suggest
$\gamma$ could be driven from any ''initial'' value to unity by the
cosmological expansion (more precisely, ST theories are driven to GR, as
soon as these theories fulfil some (not very constraining) conditions), and
also that the $1-\gamma$ current value could be of order $10^{-7}$ or $10^{-8}$ \cite{DNPRL93,DNPRD93}. On the other hand, $c^{-4}$ terms are
typically $10^{6}$ to $10^{8}$\ smaller than $c^{-2}$ terms. Hence, as much as $c^{-4}$ space-space metric terms have to be taken into account in light propagation
problems, it is necessary to include also a possible divergence from GR in
the proposed metric, at least in the $c^{-2}$ terms, for numerical coherence. Hence it is required to
upgrade the (extended version \cite{MCPRD09} of) IAU2000 metric to the ST
case. 

In principle, this would call for new definitions of multipolar moments at the $c^{-2}$ level. But other publications porposed  $c^{-2}$ mutlipolar moments in ST theories \cite{KVPR04} or in a parametrized post-newtonian framework \cite{KSPrD00}. Thus we do not discuss this point in this paper and focus only on the $c^{-4}$ metric side of the problem. 

In section \ref{sec:term}, one defines a terminology relevant to the considered problem.
Since the Einstein conformal representation plays a central role in the
approach followed in this paper, the section \ref{sec:EivsJo} is devoted to the conformal
link between the representations of the ST theories, and the related
notations we will use. Section \ref{sec:TFE} is dedicated to the derivation of the ST
field equations up to $O\left( c^{-5}\right) $ terms. In section \ref{sec:omega0}, these
field equations are rewritten up to $O\left( c^{-5},\omega
_{0}^{-1}c^{-4}\right) $ terms, for applications taking explicitely our
present knowledge on ST theories into account. Finally, considering
applications to Solar System-like systems, one defines strongly hierarchized
system in section \ref{sec:hierarchy}.\ This is made by defining a quantity $\mu $ that quantifies how much
the system is gravitationally dominated by its most massive body.
Accordingly, the field equations are rewritten up to $O\left( c^{-5},\omega
_{0}^{-1}c^{-4},\mu c^{-4}\right) $ terms. In this case, the explicit solution that can
be used in relevant applications is written in well-suited coordinates.

%______________________________________________________________________
\section{Terminology : definition of the PN/BM and PN/RM metrics}
\label{sec:term}

The PN approximation is based on the assumption of a weak gravitational
field and weak velocities for both the sources and the
(test) body (ie. velocities of the order $\sqrt{GM/r}$ or less, $M$
being some caracteristic mass of the system). It formally consists in looking for solutions under the form of
an expansion in powers of $1/c$. The usually so-called $n$PN order terms in the metric, leading to $%
c^{-2n}$ terms in the equation of motion of a body describing a bounded orbit, are terms
of orders $c^{-2n-2}$ in $g_{00}$, $c^{-2n-1}$ in $g_{0i}$ and $c^{-2n}$ in $%
g_{ij}$. In this case, the Ricci tensor components have to be developped the same way as (18) in \cite{MCPRD09}. In this paper, a metric developped this way will be refered as the $n$PN/BM metric (BM meaning "Bounded Motion" \ for test particles). It is particularly well-adapted for
studying bounded motions in systems made by non-relativistic massive bodies, as the
Solar System is. 

However, since we are interested in the propagation of light, we are lead to relax the hypothesis on the velocity of the test particle whose motion is considered. Of course, this doesn't change the metric, but the terms to be
considered in the metric components are not the same as in the PN/BM
problem.\ Indeed, the terms leading to $c^{-2n}$ terms in the equation of
motion of a test particle moving with relativistic velocity  are terms of
order $c^{-2n}$ in all the components $g_{\alpha \beta }$ (ie. in both $g_{00}$, $g_{0i}$
and $g_{ij}$). In this case, the Ricci tensor components have to be developped the same way as (19) in \cite{MCPRD09}. In this paper, a metric developped this way will be refered as the $n$PN/RM
metric (RM meaning "Relativistic Motion" \ for test particles). A PN/RM
metric is particularly well-suited for studying \textit{relativistic} motions of test bodies (for instance, the propagation of light) in systems made by non-relativistic bodies, as the Solar System is. 

\begin{figure}
\includegraphics[scale=0.25]{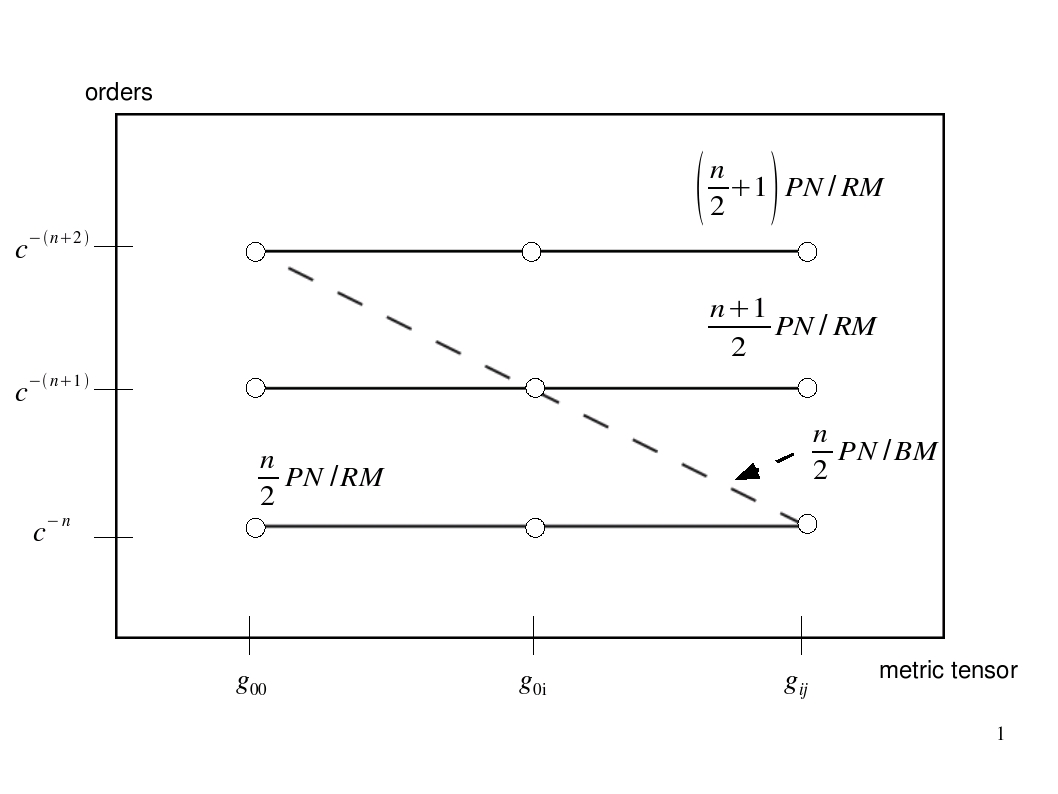}
\caption{General scheme of orders taken into account in PN/BM and PN/RM metrics}
\label{fig:PMvsPN62}
\end{figure}

PN/BM and PN/RM orders are illustrated in figure \ref{fig:PMvsPN62}.

The present paper deals with the PN/RM problem since we are concerned in propagation of light.

%______________________________________________________________________
\section{The ST theories in Einstein vs Jordan representations}
\label{sec:EivsJo}

The Jordan representation of the ST theory is described by the action 
\begin{eqnarray}
S&=&\frac{c^{4}}{16\pi G}\int d^{4}x\sqrt{-g}\left[ \Phi R-\frac{\omega \left(
\Phi \right) }{\Phi }g^{\alpha \beta }\partial _{\alpha }\Phi \partial
_{\beta }\Phi \right]  \notag \\
&+&\int d^{4}x\sqrt{-g}L_{NG}\left( \Psi ,g_{\mu \nu
}\right) .  \label{jrepr}
\end{eqnarray}
In this representation, the gravitational sector of the theory is described
by the Jordan metric $g_{\alpha \beta }$ and the scalar field $\Phi $, while
the non-gravitational fields are symbolically represented by $\Psi $. The
scalar field couples directly with the metric, leading to rather complicated
field equations. Besides, the kinetic term associated to the scalar field
doesn't have the standard form, involving a scalar field function $\omega $,
the function characterizing the ST theory we are dealing with. On the other
hand, the non-gravitational lagrangian $L_{NG}$ doesn't depend on the scalar
field, leading to simple equations of motion ($\nabla _{\alpha }T^{\alpha
\beta }=0$), with the nice consequence that the weak equivalence principle
applies in this representation of the theory.

To overcome the just mentionned drawbacks, one could be tempted to resort to
a dependent variables change 
\begin{equation*}
\left( g_{\alpha \beta },\Phi \right) \longrightarrow \left( \overset{\_}{g}%
_{\alpha \beta },\varphi \right)
\end{equation*}
chosen in such a way that the action (\ref{jrepr}) transforms into 
\begin{eqnarray}
S&=&\frac{c^{4}}{16\pi G}\int d^{4}x\sqrt{-\overset{\_}{g}}\left[ \overset{\_}{%
R}-2\overset{\_}{g}^{\alpha \beta }\partial _{\alpha }\varphi \partial
_{\beta }\varphi \right] \notag \\
&+&\int d^{4}x\sqrt{-\overset{\_}{g}}\overset{\_}{L}_{NG}  \label{erepr}
\end{eqnarray}
$\overset{\_}{L}_{NG}$ depending on $\Psi $, $\overset{\_}{g}_{\mu \nu }$
and $\varphi $ in a way to be precised later ($\overset{\_}{R}$ and $%
\overset{\_}{g}$\ correspond to $R$\ and $g$, but with $g_{\alpha \beta }$\
replaced by $\overset{\_}{g}_{\alpha \beta }$). Since the scalar field
doesn't couple with the metric, (\ref{erepr}) is refered to as the Einstein
representation of the theory. 

The form (\ref{erepr}) is achieved by
considering a conformal transformation of the metric 
\begin{equation}
g_{\alpha \beta }=A\left( \varphi \right) ^{2}\overset{\_}{g}_{\alpha \beta
}.  \label{transfoconf}
\end{equation}
From the induced transformation of the Ricci scalar \cite{Wbook84}, and up to
a divergence term, (\ref{erepr}) turns into 
\begin{eqnarray}
S&=&\frac{c^{4}}{16\pi G}\int d^{4}x\sqrt{-g}[ \frac{1}{A^{2}}R \notag\\
&&~~~~~+\left\{ \frac{6}{A^{4}}-\frac{2}{A^{2}}\left( \frac{d\varphi }{dA}\right)
^{2}\right\} g^{\alpha \beta }\partial _{\alpha }A\partial _{\beta }A] \notag \\
&+&\int d^{4}x\sqrt{-g}A^{-4}\overset{\_}{L}_{NG}.  \label{jfrome}
\end{eqnarray}
Comparing (\ref{jfrome}) with (\ref{jrepr}) suggests :

- the link between the Jordan and Einstein representations of the scalar
field 
\begin{equation}
\Phi =\frac{1}{A\left( \varphi \right) ^{2}};  \label{Adef}
\end{equation}

- the identification 
\begin{equation}
\frac{\omega \left( \Phi \right) }{\Phi }g^{\alpha \beta }\partial _{\alpha
}\Phi \partial _{\beta }\Phi =\left\{ \frac{2}{A^{2}}\left( \frac{d\varphi }{%
dA}\right) ^{2}-\frac{6}{A^{4}}\right\} g^{\alpha \beta }\partial _{\alpha
}A\partial _{\beta }A;  \label{stfctslink}
\end{equation}

- the Einstein representation of the non gravitational lagrangian 
\begin{eqnarray*}
\overset{\_}{L}_{NG}&=&A\left( \varphi \right) ^{4}L_{NG}\left( \Psi ,g_{\mu
\nu }\right) \\
&=&A\left( \varphi \right) ^{4}L_{NG}\left( \Psi ,A\left( \varphi
\right) ^{2}\overset{\_}{g}_{\mu \nu }\right) .
\end{eqnarray*}
(\ref{Adef}) and (\ref{stfctslink}) lead to the link between the functions $\omega \left(
\Phi \right) $ and $A\left( \varphi \right) $ (equivalently characterizing
the considered ST theory) 
\begin{equation}
\left( 2\omega +3\right) \left( \frac{d\ln \left| A\right| }{d\varphi }%
\right) ^{2}=1.  \label{sclink}
\end{equation}
(\ref{sclink}) requires $\omega >-3/2$.\ This results from the fact one has
imposed the sign of the scalar kinetic energy term in Einstein
representation (\ref{erepr}) in order to ensure the dynamical stability of
the theory \cite{WLNP07}.

The link between the stress tensor components in the two representations
follows from the general stress tensor definition 
\begin{eqnarray*}
\delta g^{\alpha \beta } &\longrightarrow &\delta
\int d^{4}x\sqrt{-g}L_{NG}\equiv -\frac{1}{2}\int d^{4}x\sqrt{-g}T_{\alpha
\beta }\delta g^{\alpha \beta } \\
\delta \overset{\_}{g}^{\alpha \beta } &\longrightarrow &\delta \int d^{4}x\sqrt{-\overset{\_}{g}}\overset{\_}{L}_{NG}\equiv -%
\frac{1}{2}\int d^{4}x\sqrt{-\overset{\_}{g}}\overset{\_}{T}_{\alpha \beta
}\delta \overset{\_}{g}^{\alpha \beta }.
\end{eqnarray*}
Since $\sqrt{-\overset{\_}{g}}\overset{\_}{L}_{NG}=\sqrt{-g}L_{NG}$, and
since the scalar field is not varied in this metric variation process (no ambiguity since the
two versions $\varphi $ and $\Phi $\ of the scalar field are related in the
non metric dependent way (\ref{Adef})), it directly turns that 
\begin{equation*}
\overset{\_}{T}_{\alpha \beta }=A^{2}T_{\alpha \beta }.
\end{equation*}
For the mixed and contravariant components, it follows 
\begin{equation*}
\overset{\_}{T}_{\alpha }^{\beta }=A^{4}T_{\alpha }^{\beta }\text{ \ \ ( }%
\Longrightarrow \;\overset{\_}{T}=A^{4}T\text{ ), \ \ }\overset{\_}{T}%
^{\alpha \beta }=A^{6}T^{\alpha \beta }
\end{equation*}
the indexes being raised/lowered by the metric involved in the corresponding
representation.

\bigskip

\textit{Eliminating }$A$

\bigskip

It is clear $A$ can be eliminated between the two representations of the
scalar field using (\ref{Adef}).\ This way, the considered ST theory is
represented by the function $\Phi \left( \varphi \right) $ in Einstein
representation. From (\ref{sclink}), this function is linked to $\omega
\left( \Phi \right) $ by 
\begin{equation}
\left( 2\omega +3\right) \left( \frac{d\ln \Phi }{d\varphi }\right)
^{2}=4.  \label{sclink2}
\end{equation}
The conformal transformation (\ref{transfoconf}), the link between the non
gravitational lagrangian and stress tensor representations now write 
\begin{equation}
\overset{\_}{g}_{\alpha \beta }=\Phi  g_{\alpha \beta }
\label{transfoconf2}
\end{equation}
\begin{equation*}
\overset{\_}{L}_{NG}=\Phi ^{-2}L_{NG}\left( \Psi ,\Phi ^{-1}\overset{\_}{g}%
_{\mu \nu }\right)
\end{equation*}
\begin{eqnarray}
\overset{\_}{T}_{\alpha \beta }&=&\Phi ^{-1}T_{\alpha \beta },\notag\\
\overset{\_}{T}_{\alpha }^{\beta }&=&\Phi ^{-2}T_{\alpha }^{\beta }(\Longrightarrow \;\overset{\_}{T}=\Phi ^{-2}T) \notag\\
\overset{\_}{T}^{\alpha \beta }&=&\Phi ^{-3}T^{\alpha \beta }.  \label{transfostress}
\end{eqnarray}

As it turns from equation (\ref{sclink2}), $\varphi $ is defined up to
a sign and an additive constant.\ When needed in the following, the sign
will be fixed by the choice 
\begin{equation}
\sqrt{2\omega +3}\frac{d\ln \Phi }{d\varphi }=2.  \label{sclink3}
\end{equation}
%______________________________________________________________________
\section{The field equations}
\label{sec:TFE}

%____________________________________________
\subsection{Using the Einstein representation gravitational
field variables}

From (\ref{erepr}), the field equations can be written 
\begin{eqnarray}
\overset{\_}{R}^{\alpha \beta } &=&\frac{8\pi G}{c^{4}}\left( \overset{\_}{T}%
^{\alpha \beta }-\frac{1}{2}\overset{\_}{T}\overset{\_}{g}^{\alpha \beta
}\right) +2\overset{\_}{\partial }^{\alpha }\varphi \overset{\_}{\partial }%
^{\beta }\varphi   \notag \\
\partial _{\alpha }\left( \sqrt{-\overset{\_}{g}}\overset{\_}{\partial }%
^{\alpha }\varphi \right) &=&\frac{2\pi G}{c^{4}}\overset{\_}{T}\sqrt{-%
\overset{\_}{g}}\frac{d\ln \Phi }{d\varphi } \label{fieldeq}
\end{eqnarray}
the function $\Phi \left( \varphi \right) $\ characterizing the ST theory
explicitly entering the scalar field equation. As usual in PN
approximation, let us write the scalar field as 
\begin{equation}
\varphi =\varphi _{0}+\frac{\overset{\left( 2\right) }{\varphi }}{c^{2}}+%
\frac{\overset{\left( 4\right) }{\varphi }}{c^{4}}+O\left( c^{-5}\right)
\label{escdevel}
\end{equation}
where $\varphi _{0}$\ is constant and $\overset{\left( 2\right) }{\varphi }$ and $\overset{\left( 4\right) }{\varphi }$ are zeroth order terms. (Remark that, since $\varphi $ is defined
up to an additive constant, it is not restrictive to set $\varphi _{0}=0$.) 
As a consequence, it turns out that, under the standard PN assumptions, $\overset{\_}{R}^{ij}=O\left( c^{-4}\right) $, so that the Strong Spatial Isotropy Condition (SSIC) \cite{DSXPRD91} applies in this representation.\ It is then possible to choose a coordinate system in which
the Einstein metric takes the following form at the 2PN/RM level \cite{MCPRD09}
\begin{eqnarray}
\overset{\_}{g}_{00} &=&-1+\frac{2w}{c^{2}}-\frac{2w^{2}}{c^{4}}+O\left(
c^{-5}\right)  \label{emetric} \\
\overset{\_}{g}_{0i} &=&-\frac{4w_{i}}{c^{3}}+O\left( c^{-5}\right)  \notag
\\
\overset{\_}{g}_{ij} &=&\delta _{ij}\left( 1+\frac{2w}{c^{2}}+\frac{2w^{2}}{%
c^{4}}\right) +\frac{4\tau _{ij}}{c^{4}}+O\left( c^{-5}\right) .  \notag
\end{eqnarray}
The scalar field function develops as 
\begin{equation}
\Phi \left( \varphi \right) =1+\frac{1}{c^{2}}\Phi _{0}^{\prime }\overset{%
\left( 2\right) }{\varphi }+\frac{1}{c^{4}}\left( \Phi _{0}^{\prime }%
\overset{\left( 4\right) }{\varphi }+\frac{1}{2}\Phi _{0}^{\prime \prime }%
\overset{\left( 2\right) }{\varphi }^{2}\right) +O\left( c^{-5}\right)
\label{jscdevel}
\end{equation}
where $\Phi _{0}$ ($=1$), $\Phi _{0}^{\prime }$ and $\Phi _{0}^{\prime
\prime }$\ stand for the values of $\Phi $\ and its derivatives at $\varphi
_{0}$. Putting $\Phi _{0}=1$ is not restrictive since $\Phi $ enters (\ref
{fieldeq}) throught its logarithm derivative.\ Setting 
\begin{eqnarray}
\sigma &=&\frac{1}{c^{2}}\left( T^{00}+T^{kk}\right)  \label{sources1} \\
\sigma ^{i} &=&\frac{1}{c}T^{0i}  \notag \\
\sigma ^{ij} &=&T^{ij}-T^{kk}\delta _{ij}\text{ \ \ \ \ \ ( }\Longrightarrow
\;\sigma ^{kk}=-2T^{kk}\text{\ )}  \notag
\end{eqnarray}
(which, from standard PN asumptions, are $c^{0}$ order quantities) and
using (\ref{transfostress}), the $\left( 00\right) $, $\left( 0i\right) $
and $\left( ij\right) $\ field equations (\ref{fieldeq}) lead respectively
to 
\begin{equation}
\triangle w+\frac{1}{c^{2}}\left( 3\partial _{tt}w+4\partial
_{tk}w_{k}\right) +\frac{3}{c^{2}}\Phi _{0}^{\prime }\overset{\left(
2\right) }{\varphi }\triangle w=-4\pi G\sigma +O\left( c^{-3}\right)
\label{eg00}
\end{equation}
\begin{equation}
\triangle w_{i}-\partial _{ik}w_{k}-\partial _{ti}w=-4\pi G\sigma
^{i}+O\left( c^{-2}\right)  \label{eg0i}
\end{equation}
\begin{eqnarray}
\Theta _{ij}\left( \tau _{kl}\right) &=&\partial _{i}w\partial
_{j}w-\partial _{t}\left( \partial _{i}w_{j}+\partial _{j}w_{i}\right)\notag \\
&-&2\delta _{ij}\partial _{t}\left( \partial _{t}w+\partial _{k}w_{k}\right)
\label{egij}\\
&&+4\pi G\sigma ^{ij}+\partial _{i}\overset{\left( 2\right) }{\varphi }%
\partial _{j}\overset{\left( 2\right) }{\varphi }+O\left( c^{-1}\right) 
 \notag \\
\end{eqnarray}
where $\Theta _{ij}$ is defined, as in \cite{MCPRD09}, by
\begin{eqnarray}
\Theta _{ij}\left( \tau _{kl}\right) &\equiv &\partial
_{ik}\tau _{jk}+\partial _{jk}\tau _{ik}-\triangle \tau _{ij}-\partial
_{ij}\tau _{kk}.  \notag
\end{eqnarray}
The scalar field equation gives 
\begin{eqnarray}
\triangle \overset{\left( 2\right) }{\varphi }&+&\frac{1}{c^{2}}\left(
-\partial _{tt}\overset{\left( 2\right) }{\varphi }+\triangle \overset{%
\left( 4\right) }{\varphi }\right) \notag\\
&+&\frac{1}{c^{2}}\left( 4\Phi _{0}^{\prime
}-\frac{\Phi _{0}^{\prime \prime }}{\Phi _{0}^{\prime }}\right) \overset{%
\left( 2\right) }{\varphi }\triangle \overset{\left( 2\right) }{\varphi } \notag\\
&=&-2\pi G\Phi _{0}^{\prime }\left( \sigma +\frac{\sigma ^{kk}}{c^{2}}\right)
+O\left( c^{-3}\right) .  \label{esc}
\end{eqnarray}
Remark that, in contrast to the GR case, the (00) equation is not linear,
because of the $c^{-2}\overset{\left( 2\right) }{\varphi }\triangle w$\
term. The scalar field equation also contains a non-linear $c^{-2}$\ term.

Now, combining (\ref{eg00}) and (\ref{esc}) leads to 
\begin{equation*}
\triangle \left( \overset{\left( 2\right) }{\varphi }-\frac{1}{2}\Phi
_{0}^{\prime }w\right) =O\left( c^{-2}\right) .
\end{equation*}
Accordingly, let us put 
\begin{equation}
\overset{\left( 2\right) }{\varphi }=\frac{1}{2}\Phi _{0}^{\prime }w.
\label{solphi2}
\end{equation}
Hence, defining $\chi \equiv \overset{\left( 4\right) }{\varphi }/\Phi
_{0}^{\prime }$ 
\begin{equation}
\varphi =\varphi _{0}+\frac{\Phi _{0}^{\prime }}{2c^{2}}w+\frac{\Phi
_{0}^{\prime }}{c^{4}}\chi +O\left( c^{-5}\right) .  \label{escdevel2}
\end{equation}
The metric field variables $w$, $w_{i}$ and $\tau _{ij}$ are now decoupled
from the scalar (remaining) one $\chi $. The system constraining $w$, $w_{i}$
and $\tau _{ij}$ now writes 
\begin{equation}
\triangle w+\frac{1}{c^{2}}\left( 3\partial _{tt}w+4\partial
_{tk}w_{k}\right) +\frac{3}{2c^{2}}\Phi _{0}^{\prime 2}w\triangle w=-4\pi
G\sigma +O\left( c^{-3}\right)  \label{eg00a}
\end{equation}
\begin{equation}
\triangle w_{i}-\partial _{ik}w_{k}-\partial _{ti}w=-4\pi G\sigma
^{i}+O\left( c^{-2}\right)  \label{eg0ia}
\end{equation}
\begin{eqnarray}
\Theta _{ij}\left( \tau _{kl}\right) &=&-\partial _{ti}w_{j}-\partial
_{tj}w_{i}\label{egija}\\
&+&\left( 1+\frac{1}{4}\Phi _{0}^{\prime 2}\right) \partial
_{i}w\partial _{j}w\notag\\
&-&2\delta _{ij}\left( \partial _{tt}w+\partial
_{tk}w_{k}\right) +4\pi G\sigma ^{ij}+O\left( c^{-1}\right)  \notag
\end{eqnarray}
$\chi $ being obtained in a second step, by solving 
\begin{eqnarray}
\triangle \chi &-&2\left( \partial _{tt}w+\partial _{tk}w_{k}\right)\label{esca} \\
&+&\frac{1}{%
4}\left( \Phi _{0}^{\prime 2}-\Phi _{0}^{\prime \prime }\right) w\triangle
w=-2\pi G\sigma ^{kk}+O\left( c^{-1}\right) .   \notag
\end{eqnarray}
Remark that $\triangle w$\ may be replaced by $-4\pi G\sigma $\ in the non
linear terms of equations (\ref{eg00a}) and (\ref{esca}).

%____________________________________________
\subsection{Back to Jordan representation}

Let us use the function $\omega \left( \Phi \right) $ and its derivative $%
\omega ^{\prime }\left( \Phi \right) $\ instead of $\Phi ^{\prime }\left(
\varphi \right) $ and $\Phi ^{\prime \prime }\left( \varphi \right) $. One
finds, using (\ref{sclink3})

\begin{eqnarray*}
\Phi _{0}^{\prime } &=&\frac{2}{\sqrt{2\omega _{0}+3}} \\
\Phi _{0}^{\prime \prime } &=&\frac{4}{2\omega _{0}+3}\left( 1-\frac{\omega
_{0}^{\prime }}{2\omega _{0}+3}\right) =\Phi _{0}^{\prime 2}-\frac{4\omega
_{0}^{\prime }}{\left( 2\omega _{0}+3\right) ^{2}}.
\end{eqnarray*}
One now goes back to Jordan representation using (\ref{transfoconf2}), with,
from (\ref{jscdevel}) and (\ref{escdevel2}),
\begin{eqnarray}
\Phi ^{-1}&=&1-\frac{2w}{c^{2}\left( 2\omega _{0}+3\right) }\label{eq:phim1}\\
&+&\frac{1}{%
c^{4}\left( 2\omega _{0}+3\right) }\left[ \frac{2}{2\omega _{0}+3}\left( 1+%
\frac{\omega _{0}^{\prime }}{2\omega _{0}+3}\right) w^{2}-4\chi \right] \notag\\
&+&O\left( c^{-5}\right) . \notag
\end{eqnarray}
Now let us put 
\begin{eqnarray*}
\gamma &=&\frac{\omega _{0}+1}{\omega _{0}+2}\\
\beta &=&1+\frac{\omega _{0}^{\prime }}{\left( 2\omega _{0}+3\right) \left( 2\omega
_{0}+4\right) ^{2}}\\
G_{eff}&=&\frac{2\omega _{0}+4}{2\omega_{0}+3}G
\end{eqnarray*}
and let us define 
\begin{equation*}
\left( U,U_{i},U_{ij},P\right) =\frac{2\omega _{0}+4}{2\omega _{0}+3}\left(
w,w_{i},\tau _{ij},\chi \right)
\end{equation*}
and the related quantities $\left( W,W_{i},W_{ij}\right) $ by 
\begin{eqnarray*}
W&=&U+\left( 1-\gamma \right) \frac{P}{c^{2}}\\
W_{i}&=&U_{i}\\
W_{ij}&=&U_{ij}-\left( 1-\gamma \right) P\delta _{ij}.
\end{eqnarray*}
Using (\ref{emetric}) and (\ref{eq:phim1}), one gets the Jordan metric 
\begin{eqnarray}
g_{00} &=&-1+\frac{2W}{c^{2}}-\beta \frac{2W^{2}}{c^{4}}+O\left(
c^{-5}\right)  \label{jmetric} \\
g_{0i} &=&-\left( \gamma +1\right) \frac{2W_{i}}{c^{3}}+O\left( c^{-5}\right)
\notag \\
g_{ij} &=&\delta _{ij}\left\{ 1+\gamma \frac{2W}{c^{2}}+\left( \gamma
^{2}+\beta -1\right) \frac{2W^{2}}{c^{4}}\right\}\notag \\
 &+&\left( \gamma +1\right) 
\frac{2W_{ij}}{c^{4}}+O\left( c^{-5}\right) .  \notag
\end{eqnarray}
where the functions $\left( W,W_{i},W_{ij},P\right) $\ satisfy the following field equations -- after some algebra from (\ref
{eg00a}-\ref{esca}) and (\ref{sources1}) 
\begin{eqnarray}
&&\square W+\frac{1+2\beta -3\gamma }{c^{2}}W\triangle W+\frac{2}{c^{2}}\left(
1+\gamma \right) \partial _{t}J \notag\\
&&~~~~=-4\pi G_{eff}\Sigma +O\left( c^{-3}\right)\notag \\
&&\triangle W_{i}-\partial _{i}J =-4\pi G_{eff}\Sigma ^{i}+O\left(
c^{-2}\right)  \notag \\
&&\triangle W_{ij}+\partial _{i}W\partial _{j}W+2\left( 1-\beta \right) \delta
_{ij}W\triangle W\notag \\
&&~~-\partial _{i}J_{j}-\partial _{j}J_{i}-2\gamma \delta
_{ij}\partial _{t}J=-4\pi G_{eff}\Sigma ^{ij}+O\left( c^{-1}\right) 
\notag \\
&&\triangle P+2\frac{\beta -1}{1-\gamma }W\triangle W-2\partial _{t}J  \notag \\
&&~~~~=-4\pi G_{eff}\frac{\Sigma ^{kk}}{3\gamma -1}+O\left( c^{-1}\right) . \label{fieldeq2} 
\end{eqnarray}
In (\ref{fieldeq2}), one has set 
\begin{eqnarray}
J &=&\partial _{t}U+\partial _{k}U_{k} \label{eq:jaugeJ} \\
&=&\partial _{t}W+\partial _{k}W_{k}+O\left( c^{-2}\right) \notag \\
J_{i} &=&\partial _{k}U_{ik}-\frac{1}{2}\partial _{i}U_{kk}+\partial
_{t}U_{i} \notag\\
&=&\partial _{k}W_{ik}-\frac{1}{2}\partial _{i}W_{kk}+\partial _{t}W_{i}-%
\frac{1-\gamma }{2}\partial _{i}P \label{eq:ji} 
\end{eqnarray}
and, for the matter part of the equations
\begin{eqnarray*}
&&\Sigma =\frac{1}{c^{2}}\left( T^{00}+\gamma T^{kk}\right) \\
&&\Sigma ^{i} =\frac{1}{c}T^{0i} \\
&&\Sigma ^{ij} =T^{ij}-\gamma T^{kk}\delta _{ij}\text{ \ \ \  (}%
\Longrightarrow \;\Sigma ^{kk}=-\left( 3\gamma -1\right) T^{kk}\text{)}
\end{eqnarray*}
Let us remark that the quantity $2\left( \beta -1\right) /\left( 1-\gamma
\right) $ ($=\omega _{0}^{\prime }\left( 2\omega _{0}+3\right) ^{-1}\left(
2\omega _{0}+4\right) ^{-1}$) is not diverging, even if $\gamma $\ is
(arbitrarily) close to unity. Besides, no new PN parameter appears neither in the $c^{-4}$ space-space part of the metric nor in the corresponding field equations, as stressed in \cite{DFPRD96}. 

This form is relevant in all sufficiently weak gravitational field, even in
systems where the ST theory is not very close to GR, i.e. the PN parameters $%
\gamma $ and $\beta $ are not necessarily very close to unity. A priori, this may occur even if $\gamma $ and $\beta $ are close to unity in some (other)
regions of the universe, as in the Solar System, as soon as the ST theory is
not (in some sense) close to the Brans-Dicke one (in Brans-Dicke gravity, $\omega $ doesn't depend
on the scalar field, so that it as the same value in all the space-time regions of the universe).

%______________________________________________________________________
\subsection{Harmonic gauges}

Since the use of the harmonic gauge (HG) is recommended by the IAU, let us
consider the field equations in this gauge.\ Of course one has to
specify the representation in which the HG is prescribed. The Jordan HG
condition reads 
\begin{equation*}
g^{\alpha \beta }\Gamma _{\alpha \beta }^{\sigma }=0
\end{equation*}
and it leads to, for the space ($\sigma =k$) component 
\begin{equation}
\left( \gamma -1\right) \partial _{k}U=O\left( c^{-2}\right) .
\label{STJHGk}
\end{equation}
As expected from known results in GR \cite{DSXPRD91}, this condition reduces to a triviality in the case $\gamma
=1 $.\ On the other hand, (\ref{STJHGk}) with $\gamma \neq 1$
shows that the coordinate system in which the metric takes the (Jordan) form
(\ref{jmetric}), corresponding to SSIC in Einstein representation, doesn't
encompass (Jordan) harmonic coordinates in the ST case. In other terms,
(Jordan) harmonic coordinates are incompatible with the SSIC in Einstein
representation.

One could rather choose to impose the HG condition on the metric in Einstein
representation 
\begin{equation*}
\overset{\_}{g}^{\alpha \beta }\overset{\_}{\Gamma }_{\alpha \beta }^{\sigma
}=0
\end{equation*}
since the Einstein metric (\ref{emetric})\ satisfies the SSIC. That means
one imposes $w$, $w_{i}$ and $\tau _{ij}$ to satisfy, in addition to the
field equations, the relations 
\begin{eqnarray*}
\partial _{t}w+\partial _{k}w_{k} &=&O\left( c^{-2}\right) \\
\partial _{k}\tau _{ik}-\frac{1}{2}\partial _{i}\tau _{kk}+\partial
_{t}w_{i} &=&O\left( c^{-1}\right) .
\end{eqnarray*}
Translated in terms of $\left( U,U_{i},U_{ij}\right) $, this takes exactly
the same form, i.e., using (\ref{eq:jaugeJ}-\ref{eq:ji})
\begin{equation}
J=O\left( c^{-2}\right) \text{ \ \ , \ \ }J_{i}=O\left( c^{-1}\right) .
\label{eharmo}
\end{equation}
It turns out this corresponds to the Nutku gauge constraints (\cite{KVPR04},\cite{XNDetASR09}), meaning that imposing the HG in the Einstein representation is equivalent to impose the Nutku gauge in the Jordan representation.
Using (\ref{eharmo}), the three first equations of (\ref{fieldeq2}) take the reduced form 
\begin{eqnarray}
&&\square W+\frac{1+2\beta -3\gamma }{c^{2}}W\triangle W \notag \\
&&~~~~=-4\pi G_{eff}\Sigma +O\left( c^{-3}\right) \label{eqf1} \\
&&\triangle W_{i} =-4\pi G_{eff}\Sigma ^{i}+O\left( c^{-2}\right) \label{eqf2} \\
&&\triangle W_{ij}+\partial _{i}W\partial _{j}W+2\left( 1-\beta \right) \delta
_{ij}W\triangle W \notag \\
&&~~~~=-4\pi G_{eff}\Sigma ^{ij}+O\left( c^{-1}\right) \label{eqf3}
\end{eqnarray}
while the equation corresponding to the scalar degree of freedom and the harmonic constraints read
\begin{eqnarray}
&&\triangle P+2\frac{\beta -1}{1-\gamma }W\triangle W =-4\pi G_{eff}\frac{%
\Sigma ^{kk}}{3\gamma -1}+O\left( c^{-1}\right)\\
&&\partial _{t}W+\partial _{k}W_{k}=O\left( c^{-2}\right)\\
&&\partial _{k}W_{ik}-\frac{1}{2}\partial _{i}W_{kk}+\partial _{t}W_{i}-%
\frac{1-\gamma }{2}\partial _{i}P =O\left( c^{-1}\right) \label{eq:Ji0}
\end{eqnarray}
or equivalently (after elimination of the scalar field $P$)
\begin{eqnarray}
&&\partial _{t}W+\partial _{k}W_{k}=O\left( c^{-2}\right) \label{eqf4}\\
&&\partial _{ik}W_{ik}-\frac{1}{2}\triangle W_{kk}+\partial _{ti}W_{i}+\left(
\beta -1\right) W\triangle W \notag \\
&&~~~~=-2\pi G_{eff}\frac{1-\gamma }{3\gamma -1}%
\Sigma ^{kk}+O\left( c^{-1}\right)  \label{eqf5}\\
&&\partial _{ik}W_{jk}+\partial _{ti}W_{j} =\partial _{jk}W_{ik}+\partial
_{tj}W_{i}+O\left( c^{-1}\right) . \label{eqf6}
\end{eqnarray}
The last equation refers to the fact that $\partial _{i}P$ (given by (\ref{eq:Ji0})) is
a gradient.

One could note these equations are coherent with 1.5PN/BM equations assumed in \cite{KSPrD00}.

%______________________________________________________________________
\section{Simplified forms that can be used considering present constraints on gravity}
\label{sec:omega0}

In the inner solar system, the gravitational field is such that 
\begin{equation*}
\frac{2U}{c^{2}}\sim 10^{-6}\text{ to }10^{-8}.
\end{equation*}
On the other hand, from experimental/observational constraints \cite{WLivRev06}
\begin{equation*}
\left| \gamma -1\right| \lesssim 10^{-5}\text{ ie. }\omega _{0}\gtrsim 10^{5}.
\end{equation*}
This means $ \gamma -1 $ (or $\omega _{0}^{-1}$) could be
considered numerically as a $c^{-1}$\ (at best) order quantity.\ Hence, it is convenient to present the metric under the form of a generalized
development in both powers of $c^{-1}$ and $\omega _{0}^{-1}$.\ The useful metric resulting from (\ref{jmetric}) reads (if $\omega _{0}^{\prime }$ is not
''unreasonably large'')\ 
\begin{eqnarray*}
g_{00} &=&-1+\frac{2W}{c^{2}}-\frac{2W^{2}}{c^{4}}+O\left( c^{-5},\omega
_{0}^{-1}c^{-4}\right) \\
g_{0i} &=&-\left( \gamma +1\right) \frac{2W_{i}}{c^{3}}+O\left( c^{-5}\right)
\\
g_{ij} &=&\delta _{ij}\left[ 1+\gamma \frac{2W}{c^{2}}+\frac{2W^{2}}{c^{4}}%
\right]\\
&+&\frac{4W_{ij}}{c^{4}}+O\left( c^{-5},\omega _{0}^{-1}c^{-4}\right)
\end{eqnarray*}
where $W$, $W_{i}$ and $W_{ij}$ satisfy, from (\ref{fieldeq2})
\begin{eqnarray*}
&&\square W+\frac{4}{c^{2}}\partial _{t}J =-4\pi G_{eff}\sigma +O\left(
c^{-3},\omega _{0}^{-1}c^{-2}\right) \\
&&\triangle W_{i}-\partial _{i}J =-4\pi G_{eff}\sigma ^{i}+O\left(
c^{-2}\right) \\
&&\triangle W_{ij}+\partial _{i}W\partial _{j}W-\partial _{i}J_{j}-\partial
_{j}J_{i}-2\delta _{ij}\partial _{t}J \\
&&~~~~=-4\pi G_{eff}\sigma ^{ij}+O\left(
c^{-1},\omega _{0}^{-1}\right)
\end{eqnarray*}
and where $J_{i}$ reduces to 
\begin{equation}
\label{eq:NGiapprx}
J_{i}=\partial _{k}W_{ik}-\frac{1}{2}\partial _{i}W_{kk}+\partial
_{t}W_{i}+O\left( \omega _{0}^{-1}\right) .
\end{equation}
One remarks the field equations take exactly the same form as the GR case \cite{MCPRD09}
(with $G$ replaced by $G_{eff}$). The only remaining reference to
the scalar field is reduced to the presence of the $\gamma $ PN coefficient
in the metric tensor. Related to this, the field equation on $P$ is dropped out.

%______________________________________________________________________
\subsection{Harmonic gauge}

The corresponding harmonic equations to be used when considering known constraints on $\omega_0$ reads
\begin{eqnarray}
&&\square W =-4\pi G_{eff}\sigma +O\left( c^{-3},\omega
_{0}^{-1}c^{-2}\right) \label{eq:eqjH1} \notag\\
&&\triangle W_{i} =-4\pi G_{eff}\sigma ^{i}+O\left( c^{-2}\right) \label{eq:eqjH2} \\
&&\triangle W_{ij}+\partial _{i}W\partial _{j}W =-4\pi G_{eff}\sigma
^{ij}+O\left( c^{-1},\omega _{0}^{-1}\right)\label{eq:eqjH3} \notag
\end{eqnarray}

with gauge conditions 

\begin{eqnarray}
&&\partial _{t}W+\partial _{k}W_{k}=O\left( c^{-2}\right) \label{equ:h0}\\
&&\partial _{k}W_{ik}-\frac{1}{2}\partial _{i}W_{kk}+\partial \label{equ:hi}
_{t}W_{i}=O\left(c^{-1}, \omega _{0}^{-1}\right).
\end{eqnarray}

%______________________________________________________________________
\section{Explicit form to be used in the case of the Solar system}
\label{sec:hierarchy}

%______________________________________________________________________
\subsection{Strongly hierarchized systems}

Let us consider the case where the system is composed by bodies of masses $%
M_{A}$. Let us consider one of these bodies, named $S$, of mass $M_{S}$.\
Let us define the parameter%
\begin{equation*}
\mu =\frac{1}{M_{S}}\sum_{A\neq S}M_{A}.
\end{equation*}%
One defines a strongly hierarchized system as a system in which it is
possible to choose the body $S$\ in such a way that%
\begin{equation*}
\mu \ll 1.
\end{equation*}
In such a system, the body S will be hereafter referred as "the
star", while the other bodies will be referred as the "planets". 

In the general relativistic $N$-body problem, multipolar moments of a body $%
A $ are defined in the coordinate system in which this body is, in some
sense, at rest. These moments are affected by coordinate transforms through a "Lorentz-like length contraction effect". These effects being of order 
$\left( u/c\right) ^{2}$, where $u$\ is the relative velocity between the
two frames, the induced effects in the metric components are of order $c^{-4}
$.

In strongly hierarchized systems, the velocity of the body $S$ is of the
order of%
\begin{equation*}
v_{S}\sim \mu v_{B}\sim \mu \sqrt{\frac{GM_{S}}{r_{S-B}}}
\end{equation*}%
where $B$ is the most
massive planet (and $r_{S-B}$\ the distance between $B$\ and the star). All the Lorentz-like contraction terms have the form
\begin{equation*}
\frac{G M_A}{r c^2} \frac{v_A^2}{c^2}.
\end{equation*}
If $A$ is a planet ($A \neq S$), this term is of order $O\left( \mu c^{-4}\right)$. If $A$ is the star ($A = S$), this term is of order $O\left( v_S^2 c^{-4}\right)$, ie. $O\left( \mu^2 c^{-4}\right)$. Hence all these terms are, at best, of order
\begin{equation*}
O\left( \mu c^{-4}\right) .
\end{equation*}

Let us also point out that, since at this level the metric depends on time through the positions of the star and the planets only, the operator $\partial _{t}$ is of order $O(\mu)$. Hence, equations (\ref{eq:eqjH1})-(\ref{equ:hi}) lead to

\begin{eqnarray}
&&\triangle W =-4\pi G_{eff}\sigma +O\left( c^{-3},\omega
_{0}^{-1}c^{-2},\mu c^{-2}\right) \label{eq:eqjH1b} \notag\\
&&\triangle W_{i} =-4\pi G_{eff}\sigma ^{i}+O\left( c^{-2},\mu c^{-1}\right) \label{eq:eqjH2b} \\
&&\triangle W_{ij}+\partial _{i}W\partial _{j}W =-4\pi G_{eff}\sigma
^{ij}+O\left( c^{-1},\omega _{0}^{-1}, \mu\right)\label{eq:eqjH3b} \notag
\end{eqnarray}

with gauge conditions 

\begin{eqnarray}
&&\partial _{t}W+\partial _{k}W_{k}=O\left( c^{-2},\mu c^{-1}\right) \label{equ:h0b}\\
&&\partial _{k}W_{ik}-\frac{1}{2}\partial _{i}W_{kk}
=O\left(c^{-1}, \omega _{0}^{-1}, \mu\right).\label{equ:hib}
\end{eqnarray}

Related to this, $\Delta_A$ defined in \cite{SKPetAJ03} leads to numerically negligible terms (see (11.4.8) in \cite{KVPR04} for the ST version). 

%______________________________________________________________________
\subsection{Application to the solar system}

In the Solar system, the most massive body $S$ is the Sun and one has
$$\mu \sim 10^{-3}.$$
Thus, it is legitimate to consider the Solar system as a strongly hierarchized system. Note that, at best, only $J_2$, $J_4$ and $J_6$ planetary terms (giant planets) could have a significant impact on laser ranging experiments at the required accuracy (see \cite{jacobson_jupiter,jacobson_saturn,jacobson_uranus,jacobson_neptune} for giant planets' multipole moments values). Hence, taking advantage that the Solar multipolar terms are very weak, the solution of the field equations (\ref{eq:eqjH2b}) -- suitable for millimetric accuracy in propagation of light -- with the harmonic constraints (\ref{equ:h0b}-\ref{equ:hib}) given in barycentric coordinates turns to be

\begin{widetext}
\begin{eqnarray}
g_{00}&=&-1+\frac{2}{c^2} \left[W_0(t,\vec{x})+W_L(t,\vec{x})\right] - \frac{2 W_S^2}{c^4}+O\left( c^{-5},\omega_{0}^{-1}c^{-4},\mu c^{-4},J^S_2 c^{-4}\right) \label{eq:hugei}\\
g_{0i}&=&- 2\frac{\gamma+1}{c^3} W^i(t,\vec{x})+O\left( c^{-5},\mu c^{-4},J^S_2 c^{-4}\right) \label{eq:wifin}\\
g_{ij}&=& \left( 1 + \frac{2 \gamma}{c^2} \left[W_0(t,\vec{x})+W_L(t,\vec{x})\right]+\frac{2 W_S^2}{c^4} \right) \delta_{ij}+4 \frac{W_{ij}}{c^4}+O\left( c^{-5},\omega_{0}^{-1}c^{-4},\mu c^{-4},J^S_2 c^{-4}\right)
\end{eqnarray}
where
\begin{eqnarray}
W_0(t,\vec{x})&=& \sum W_{A,0} \mbox{, with } W_{A,0} = G_{eff} \frac{M_A}{r_A(t,\vec{x})}\\
W_L(t,\vec{x})&=& \sum W_{A,L} \mbox{, with } W_{A,L} =- G_{eff} \sum_{n=1}^{3} M_A J_{2n}^A ~~\frac{R_A^{2n}}{r_A^{2n+1}}~~ P_{2n}\left(\frac{\hat{k}_A\cdot \vec{r}_A}{r_A} \right).
\end{eqnarray}
\begin{eqnarray}
W^i(t,\vec{x})= \sum_A W^i_A(t,\vec{x})\mbox{, with }W^i_A(t,\vec{x})&=& G_{eff} \left[- \frac{\left(\vec{r}_A \times \vec{S}_A \right)^i}{2 r_A^3}+\frac{M_A v_A^i}{r_A} \left(1+\sum_{n=1}^3 J^A_{2n} \frac{R_A^{2n}}{r_A^{2n}} P_{2n} \left(\frac{\hat{k}_A \cdot \vec{r}_A}{r_A} \right) \right) \right] \label{eq:Wifin}\\
W_S(\vec{x}) &=& G_{eff} \frac{M_S}{r_S}\\
W_{ij}(\vec{x}) &=& \frac{1}{4} \left(G_{eff} \frac{M_S}{r_S}\right)^2 \left(\frac{(x^i-x_S^i)(x^j-x_S^j)}{r_S^2} - \delta_{ij} \right) \label{eq:hugef}
\end{eqnarray}
\end{widetext}
where one has put
$$\vec{r}_A(t,\vec{x})=\vec{x}-\vec{x}_A(t) \textrm{ and } r_A(t,\vec{x})=|\vec{r}_A(t,\vec{x})|.$$
$M_A$, $r_a$, $v_A$ and $S_A$ being repectively the mass, the position and the velocity in barycentric coordinates, and the total angular momentum of the body $A$. $R_A$ and $J_n^A$ are the radius and the mass multipole coefficients of the body $A$. $P_n$ are the Legendre polynomials and $\hat{k}_A$ denotes the unit vector along the local $Z_A$ axis of each body $A$. 
The differences with the IAU2000 metric \cite{SKPetAJ03} lie in the presence of both the PN parameter $\gamma$ and the $c^{-4}$ space-space metric term. The multipolar term $W_L$ in $g_{ij}$ that has been neglected in the IAU2000 metric -- thanks to numerical considerations in the 1PN/BM case -- has to be considered here as well.

In a same way, the multipolar terms in the time-space component of the metric ($g_{0i}$) could also lead to measurable effects. Thus, one may have to consider them. 

While the rotational term in the time-space component of the metric is given as the usual Lense-Thirring term, slight modifications (spin multipoles) can in principle appear due to the differential rotation of the bodies. However, Solar seismology suggests \cite{corbard1} that the Sun's tachoclyne is at about $0.7$ Sun radius. Then, the mass concerned by the differential rotation is of order of a few percent of the total mass and thus, it might not lead to measurable effects. But, incidentally, time transfer and laser ranging experiments could suggest a way to test our knowledges on the solar interior dynamics, independently of results coming from Solar seismology. Another point coming from the solar seismology is that the solar core ($r<0.2$ Sun radius) may rotate faster than the external layers \cite{modeG1,modeG2,modeG3}. Since, the core represents a great amount of the total mass, this could affect the propagation of light at a level depending on the total angular momentum value ($\vec{S}_S$). Depending on the solar internal structure model, this may happen at the millimetric level. Using a simplified model, we show in appendix \ref{app:diffrot} how a non-rigid rotation of the Sun could also affect the metric.

Let us note that $W_L$ can be neglected for inner solar system millimetric laser ranging experiments, such as Mars laser ranging for instance.

We emphasize again that neiher the $\beta$ parameter nor the $\epsilon$ parameter are required ($\epsilon$ -- corresponding sometimes to $\Lambda$ \cite{RWPRD83} or $\delta$ \cite{lator} -- being some PN parameter often considered in the $c^{-4}$ space-space metric term \cite{ESPRD80,KZArxiv09}). This is because both the former and the latter give too small deviations from GR to be considered in $c^{-4}$ Solar system photon's trajectories calculations. This fact is known for the former from \cite{WLivRev06} and is then obvious for the latter since it is a function of $\gamma$ and $\beta$ in (non-massive-)ST theories considered here (as expressed in equation (\ref{jmetric})).

Finally, let us point that, because of the orders of magnitude in the Solar system problem, while we must know the global (ie. BCRS) 2PN/RM metric for experiments of interest -- which is given in (\ref{eq:hugei})-(\ref{eq:hugef}) -- it is not necessary to have the local (planet) 2PN/RM metrics, insofar such metrics are needed. This means that the usual local 1,5PN/BM metrics (\cite{DSXPRD91} for GR and \cite{KVPR04} for ST) are accurate enough for overlaping local coordinates charts with the global one. Hence, it is not needed to develop a complete theory of reference frames at the 2PN/RM level as long as one is interrested in millimetric laser ranging experiments in the Solar system.

%______________________________________________________________________
\section{Conclusion}

Laser ranging and time transfer in solar system could be able to reveal smoking gun arguments against GR by measuring accurately deviation from GR in well-suited experiments such as LATOR \cite{lator}, TIPO \cite{tipo}, ASTROD \cite{astrod}, ODYSSEY \cite{odyssey} or SAGAS \cite{sagas}. Such experiments will require distance measurements at millimetric level. As emphasized in \cite{MCPRD09}, $c^{-4}$ terms in $g_{ij}$ then have to be taken into account. But only monopolar terms have to be considered in the $c^{-4}$ part of the metric. Thus, definition of multipolar moments is required at the $c^{-2}$ level only. Such definitions are available for GR and TS theories \cite{DSXPRD91,KVPR04} and are in progress in the PPN formalism \cite{KSPrD00}. Moreover, since the Solar system is strongly hierarchized, only the Solar monopolar term has to be considered in the $c^{-4}$ part of the metric.

Hence, from numerical considerations that spring from present constraints on the post-Newtonian parameters and from the Solar system specificities, the 2PN/RM metric (\ref{eq:hugei})-(\ref{eq:hugef}) is sufficient for the next generation of experiments dealing with propagation of light in the Solar system.

\begin{acknowledgments}
Olivier Minazzoli wants to thank the Government of the Principality of Monaco for their financial support.
\end{acknowledgments}

\appendix
\section{Effects on the metric of a non-rigid rotation}
\label{app:diffrot}

Recent results coming from Solar seismology suggest that the Solar core rotates faster than the external layers \cite{modeG1,modeG2,modeG3}. In the following toy model we consider the mass density as spherical ($\rho(\vec{x})=\rho(r)$). As usual in Solar models \cite{corbard1}, we make the distinction between three main regions : the core region ($r \in [0,R_C \approx 0,2 R_S]$, where $R_C$ is the core radius), the radiative region ($r \in [R_C,R_R\approx 0,7 R_S]$, where $R_R$ is the tachoclyne radius) and the convection region ($r \in [R_R,R_S]$, where $R_S$ is the Sun radius). $\vec{\Omega}$ being the angular velocity, we modelize the differential rotation as follows
\begin{equation}
\vec{\Omega}(\vec{x})=\Omega(r) \hat{k}_z,
\end{equation}
with
\begin{eqnarray}
\Omega(r)=&& \Omega_C(\theta)~ \Pi \left(\frac{r}{R_c}\right) \\
&+&\Omega_R(\theta)~ \Pi \left(\frac{r-R_C}{R_R-R_C}\right) \notag \\
&+&\Omega_D(\theta)~ \Pi \left(\frac{r-R_R}{R_D-R_R}\right) \notag,
\end{eqnarray}
where 
\begin{eqnarray*}
\Pi(x) &=& 1 ~\forall~ x \in [0;1]\\
&=&0~ \mbox{ everywhere else},
\end{eqnarray*}
and where we modelize the differential rotation by a simple model in accordance with the usual model \cite{corbard2}
\begin{equation}
\Omega_B(\theta) = \Omega_B \left(1+\epsilon_B~ cos^2 \theta \right),
\end{equation}
where $\Omega_B$ and $\epsilon_B$ are constants, $B$ being any of the three previous regions ($C$, $R$ or $D$).

Solar seismology suggests that the radiative region rotates as a solid -- meaning $\Omega_R(\theta)=\Omega_R$ (ie. $\epsilon_R=0$). However, since we are interested in testing solar seismology results, let us relax this assumption. Let us write
\begin{equation}
\bar{W}^i_S(t,\vec{x})=\bar{W}^i_C(t,\vec{x})+\bar{W}^i_R(t,\vec{x})+\bar{W}^i_D(t,\vec{x}),
\end{equation}
with
\begin{equation}
\bar{W}^i_B(t,\vec{x})=G \int_B d^3X \left(\Omega_B(R) \times \vec{X} \right)^i \frac{\rho(R)}{|\vec{x}-\vec{X}|}.
\end{equation}
$\bar{W}^i_S$ being the Sun spin part of $W_i$ in (\ref{eq:wifin}). Then the solution writes
\begin{eqnarray}
&& \bar{W}^i_S(t,\vec{x}) =4 \pi G \sum_B \label{appeq:lens++}\\
&& \left(\vec{\Omega}_B \times \vec{x} \right)^i \left[\left(\frac{1}{3}+\frac{\epsilon_B}{15} \right) \frac{M_2^B}{r^3} + \frac{\epsilon_B}{35} \frac{M_4^B}{r^5} (5~cos^2 \theta-1) \right].\notag
\end{eqnarray}
Where
\begin{eqnarray*}
M_N^C &=& M_N(0,R_C),~~ M_N^R = M_N(R_C,R_R),\\
M_N^D &=& M_N(R_R,R_S),
\end{eqnarray*}
with
\begin{equation}
M_N(X,Y)=\int_X^Y R^{N+2} \rho(R) dR.
\end{equation}

First, note that a faster rotation of a rigid core will modify the value of the total angular momentum only. However, this value could be affected by a differential rotation as well.

But, one also may have to consider a term like the last term of the r.h.s. of (\ref{appeq:lens++}) in (\ref{eq:Wifin}), in order to measure possible weak effects due to differential rotations of the different stages of the Sun -- then, giving a characteristic way to put constraints on such differential rotations, which will be independent of Solar seismology and neutrino detection results. In what follows, we will refer to this term as the post-Lense-Thirring term.

As an illustration, let us consider the following toy model. Assume (1) the density decreases linearly with the distance to the center of the Sun (2) the differential rotation is independent of the distance from the center. Now consider that the photons -- used for the time transfer \cite{MCPRD09} -- graze the Sun (ie. $b=\alpha R_S$, where $b$ is the impact parameter and $\alpha(>1)$ a parameter ideally close to 1). Then the effect of the post-Lense-Thirring term is about $\alpha^{-2}\epsilon_B/11$ times the usual Lense-Thirring effect.

However, realistic models of the Sun and its rotation are expected to substantially decrease the value of this possible effect. But, this point should be verified and a specific study, that consider different realistic models of the Sun, should be done in order to clarify this issue.

\end{document}